\documentclass[runningheads]{llncs}
\usepackage{graphicx}
\usepackage{xcolor}
\usepackage{soul}
\soulregister\ref7
\soulregister\cite7

\begin{document}
\title{Quality and Reliability Metrics for IoT Systems: A Consolidated View}
\titlerunning{Quality and Reliability Metrics for IoT Systems}
%
\author{Matej Klima\inst{1}\orcidID{0000-0002-9601-8787} \and
Vaclav Rechtberger\inst{1}\orcidID{0000-0002-4127-5372} \and
Miroslav Bures\inst{1}\orcidID{0000-0002-2994-7826} \and
Xavier Bellekens\inst{2}\orcidID{0000-0003-1849-5788} \and
Hanan Hindy\inst{3}\orcidID{0000-0002-5195-8193} \and
Bestoun S. Ahmed\inst{1,4}\orcidID{0000-0001-9051-7609}}

\authorrunning{M. Klima et al.}
%
\institute{Dept. of Computer Science, FEE, Czech Technical University in Prague, Czechia
\email{\{klimama7,miroslav.bures\}@fel.cvut.cz}\\
\url{http://still.felk.cvut.cz/} \and
Dept. of Electronic and Electrical Engineering, University of Strathclyde, Glasgow, United Kingdom\\ \and
Division of Cyber Security, Abertay University, Dundee, United Kingdom\\ \and
Dept. of Mathematics and Computer Science, Karlstad University, Karlstad, Sweden}
\maketitle              
\begin{abstract}
Quality and reliability metrics play an important role in the evaluation of the state of a system during the development and testing phases, and serve as tools to optimize the testing process or to define the exit or acceptance criteria of the system. This study provides a consolidated view on the available quality and reliability metrics applicable to Internet of Things (IoT) systems, as no comprehensive study has provided such a view specific to these systems. The quality and reliability metrics categorized and discussed in this paper are divided into three categories: metrics assessing the quality of an IoT system or service, metrics for assessing the effectiveness of the testing process, and metrics that can be universally applied in both cases. In the discussion, recommendations of proper usage of discussed metrics in a testing process are then given.

\end{abstract}

\section{Introduction}

\color{blue}
Paper accepted at EAI Urb-IoT 2020 - 5th EAI International Conference on IoT in Urban Space, December 2-4, 2020.

\vspace{3mm}
https://urbaniot.eai-conferences.org/2020/
\vspace{3mm}

\color{black}

To evaluate the output quality and reliability of a System Under Test~(SUT), various characteristics and metrics are commonly used~\cite{jung2004measuring,van2013tmap,koomen2013tmap,kan2002metrics}. For example, we can give the ratio of the number of known defects in an SUT to the code lines number, the number of defects occurring in a production run of an SUT, or a ratio of time when a service provided by the SUT is available without being blocked by a defect. Thus, quality and reliability metrics can serve several purposes including, but not limited to:
(1)~monitoring the quality of a created SUT during the later development and testing phases; (2)~evaluating the effectiveness of the testing and debugging process; (3)~serving as the exit criteria between test levels and the acceptance criteria at the end of system development; (4)~evaluating the reliability of a system in its production run.

Despite the fact that the field of measurements and metrics has been adequately discussed for various aspects of software systems (for instance~\cite{chidamber1994metrics,10.1007/978-3-319-49421-0_5,chawla2016quantitative,staron2011developing,bures2015metrics,bures2015framework,dromey1995model,kan2002metrics}), no consolidated overview of the reliability and quality metrics focused on an Internet of Things (IoT) system from a high-level perspective has been published to the best of our knowledge.  

In the field of IoT quality and reliability metrics, only studies focusing on individual aspects of quality and reliability of IoT systems have been published, which will further be explored in Section~\ref{sec:related_work}.

IoT systems differ from software systems in a number of aspects, which brings specific quality assurance challenges~\cite{kiruthika2015software,marinissen2016iot,ahmed2019aspects}. As some examples, we can give (1) larger heterogeneity of used technologies, protocols and devices, creating a significantly higher number of possible configurations of a system to be tested, (2) higher demands on interoperability and flawless integration, (3) current lower level of standardization of communication protocols or (4) various privacy and security issues of the current IoT systems. Because of these differences, we consider it relevant to approach the IoT domain separately to the software domain and analyze the relevant quality and reliability metrics accordingly.

Similar consolidation work on the quality characteristics of IoT systems was recently carried out~\cite{bures2018comprehensive} by the authors and was supplemented by another recent study by White~\textit{et al}. 
Specifically, \cite{white2017quality} studied and summarized the available literature focused on quality of services~(QoS) in IoT systems. The study primarily focused on quality characteristics and the architectural perspective; however, it does not discuss particular quality metrics.

General quality characteristics differ from quality and reliability metrics by their level of detail and domain applicability.

A \textit{quality metric} provides detailed information expressed by a number and is typically defined by a formula that is based on the quantification of SUT elements, the SUT model (e.g., a number of defects on a line of code), and/or quantified information from the testing and test management process (e.g., a number of found defects).

We understand the \textit{quality characteristic} to be a general property of the SUT that can be used to carry the test planning, test strategy or test reporting; e.g., functional correctness, security, usability, or maintainability~\cite{bures2018comprehensive}. Differently to quality metrics, quality characteristics are usually not expressed by particular formulas that allow for the quantification of a measured property by a concrete number.

This paper is organized as follows. Section~\ref{sec:related_work} analyzes the existing literature related to the quality metrics of both software and IoT systems. Section~\ref{sec:metrics} provides a consolidated view on IoT-related quality and reliability metrics with references to sources originally discussing these metrics. Section~\ref{sec:discussion} discusses the consolidated overview and possible limits of this work. The last section concludes the paper.

\section{Related work}
\label{sec:related_work}

In the field of IoT systems, several studies discussing individual quality metrics have been published. These are analyzed and discussed within this section. However, these studies unequivocally lack the quality and reliability metrics applicable to IoT.

To distinguish the quality of cloud services providers, Zheng~\textit{et al.} define a quality model for cloud services CLOUDQUAL~\cite{zheng2014cloudqual}. The model consists of six different quality dimensions and metrics: availability, reliability, usability, responsiveness, security, and elasticity. Despite focusing on cloud services, some metrics are broad enough to be adapted to IoT systems.

A study by Li~\textit{et al.} proposes availability, together with currency and validity, as quality metrics for measuring the data quality in pervasive environments~\cite{li2012data}. We consider the data availability component of this work to be relevant to IoT systems.

Sollie~\cite{sollie2005security} discusses metrics for assessing the security and usability of authentication systems. From these metrics, the \textit{``Rate of User Error"} is relevant to the IoT domain.

In the recent studies by Kim~\cite{kim2016quality} and Kim~\textit{et al.}~\cite{kim2016qualityService}, quality models for the evaluation of IoT applications and services are presented. Kim discusses particular definitions of metrics and identifies four criteria: functionality, reliability, efficiency, and portability, for which various metrics are presented~\cite{kim2016quality}.

In the field of QoS measurements for IoT systems, certain quality characteristics have been categorized by Singh~\textit{et al.}~\cite{8519862}. Three main types of QoS measurements are identified: the QoS of communication, the QoS of things, and the QoS of computing. However, although the authors use the term metrics, the paper actually discusses quality characteristics (formulas defining metrics are not provided in this study). Snigdh~\textit{et al.}  published a similar categorization for wireless sensor network areas in which more metrics are discussed~\cite{snigdh2016quality}.

A comprehensive literature study of the QoS for IoT systems has been conducted by White~\textit{et al.}~\cite{white2017quality}. This study discusses three  aspects of the QoS: (1)~layers of the IoT architecture, which are the most frequent subject of QoS research; (2)~the quality factors are measured; (3)~the types of research conducted in the field. In this study, consolidated high-level quality characteristics can be found. However, no consolidated view on quality metrics with their definitions is provided, as such overview is beyond the scope of the study.

The quality of end devices in IoT systems have also been the subject of some investigation; for example, actuators in~\cite{6694051}. In this study, concrete measurements of quality parameters are presented, and the authors conclude that the main factors impacting perceived quality were \textit{``average delay"} and \textit{``packet loss"}~\cite{6694051}. These factors could also be applied to other IoT components.

One study by Staron~\textit{et al.}~\cite{10.1007/978-3-319-49421-0_5} contains metrics for measuring the quality of the system architecture, such as the number of coupled components, the number of changes in architecture per time unit, or the number of interfaces. However, the \textit{``Design Stability"} description discusses metrics that can also be applied in the IoT environment. 

Baggen~\textit{et al}. discuss a set of software code metrics impacting  maintainability~\cite{baggen2012standardized}, thereby extending the previous list of metrics proposed by Heitlager~\textit{et al.}~\cite{heitlager2007practical}. We consider these metrics to be relevant in different parts of IoT systems. The authors list the volume of the code, its redundancy, the size of its units, complexity, unit interface size, and the extent of component coupling. Although no particular definition of metrics is given in the study, they can be easily defined from these suggestions.

Besides coupling and code complexity, Pantiuchina~\textit{et al.} discuss other code quality metrics; cohesion and code readability in particular.  Formulas to compute cohesion are also provided in~\cite{pantiuchina2018improving}. The lack of code cohesion and coupling indicators is examined by Chaparro~\textit{et al.}, and detailed formulas to quantify the properties are provided in their study~\cite{chaparro2014impact}.

Code quality impacts the potential reliability and quality of an IoT system. High-quality decreases the presence of flaws in the system and positively impacts the maintainability and ease of extending the system. However, it is difficult to identify the relations between high-level quality metrics typically based on defects found in a system or the failures of the system and code quality metrics.

Defining quality metrics is generally inspired by related work focusing on more general quality characteristics. As examples of quality characteristics, we can give an overview of Sogeti’s test management approach~(TMap) methodology, which focuses on software\footnote{https://www.tmap.net/wiki/quality-characteristics} and, following the recent trends, on IoT systems\footnote{https://www.tmap.net/wiki/quality-characterstics-iot-environment}\cite{van2016iotmap}.

Regarding the security aspects, the metrics that are utilized for general-purpose networks and apply to IoT networks are extensively discussed by Hindy \textit{et al.}~\cite{9108270}. Specifically focusing on IoT security, Bonilla \textit{et al.}~\cite{8328467} proposed a particular metric for the measurement of the security level of IoT devices that we later include in our overview.

Following the literature review, it is apparent that there is a lack of comprehensive studies focusing on IoT-related quality metrics. Such a study is the subject of this paper.

\color{black}
\section{Metrics Overview}
\label{sec:metrics}
In the following section, we provide a consolidated overview of the quality and reliability metrics applicable to IoT systems. 

The scope of this study approaches the quality metrics problem from the overall view of an IoT system. Considering this scope as the delimitation, this study does not focus specifically on QoS metrics, as their goal is primarily to evaluate the performance of the network layers of such systems. Thus, the overview provided does not focus on general test coverage criteria and specific code quality metrics. We explain the reasons and provide the literature for these fields in Section~\ref{sec:discussion}.

\subsection{Methodology of this Overview}
Seven publisher databases and indexing services were used for the review: IEEE Xplore, ACM Digital Library, Springer Link, Elsevier ScienceDirect, Web of Science, Scopus, and Google Scholar. 

The generic search string (adopted in accord with the local specifics of individual databases and indexing services) is:

\textit{(`Quality Metrics' AND IoT) OR (`Quality Metrics' AND `Internet of Things') OR (`Quality Measurement' AND IoT) OR (`Quality Measurement' AND `Internet of Things') OR (`Quality Model' AND IoT) OR (`Quality Model' AND '`Internet of Things')}

where apostrophes serve to denote an exact string that must be searched for. No publication time span was set during the search.

The papers found were assessed for their relevance to the discussed topic based on the abstract and a full reading. The process was conducted using ``two pairs of eyes" verification, to minimize possible errors during the search phase. 

As a consequence of the full-text reading of relevant papers, we conducted a snowball sampling process to acquire other relevant papers discussing the topic. The quality metrics found in the papers were then consolidated.

In this study, we divide the metrics into three categories: (1)~metrics relating to the quality of an IoT system, product, or service; (2)~metrics relating to the effectiveness of the testing process of an IoT system; and (3)~metrics applicable to both previous aspects. Each of these categories is discussed in a separate subsection.


It is important to mention that the same name in the literature can refer to practically different quality metrics. For example, the availability defined by Li~\textit{et al.}~\cite{li2012data} describes the availability of data in a system; the availability defined by Zheng~\textit{et al.}~\cite{zheng2014cloudqual} describes the general availability for an IoT service. In such cases, the differences are discussed in the explanation of the metrics.

\subsection{Quality Metrics to Evaluate an IoT System or Service}

The quality and reliability of an IoT system or service can be measured in different aspects, that can be expressed by particular metrics.

\color{black}

\subsubsection{Availability.}

To measure the availability of a service or an IoT system~\cite{zheng2014cloudqual}, a metric based on the uptime ratio of the service during a specified time interval can be used:
\begin{center}
    $AV = \frac{t_{up}}{t}$, \\
\end{center}
where $t_{up}$ is the time the service was available, and $t$ is the time interval the availability was measured. The metric values range from 0 to 1, with 1 representing a 100\% availability of the service.

A similar metric can be defined for the~\textit{availability of data} provided as a part of the service as proposed by Li~\textit{et al.}~\cite{li2012data}, which can be defined as:

\begin{center}
    $DAV = 1- \frac{\sum_{i=1}^{n}max(0,  t_{i}-T^{exp})}{OP}$, \\
\end{center}

where $OP$ denotes the observation period, $n$ denotes the number of data objects received during $OP$, $t_{i}$ is the interval between the $i$th and the $i+1$th updates, and $T^{exp}$ is expiration time~\cite{li2012data}.

\subsubsection{Flaws over Time.} To measure the system reliability, a ratio of the number of critical flaws found in the system over a period during a review or after system deployment can be used: 
\begin{center}
    $FVT = \frac{n_{failure}}{n_{total}}$, \\
\end{center}

where $n_{failure}$ is the total number of failed operations and $n_{total}$ is the total number of operations that have occurred in a time interval~\cite{kim2016quality}. The metric value ranges from 0 to 1 with 0 indicating that there have been no flaws observed in the system during the measured time. 

\subsubsection{Reliability.} Alternatively, we can use an inverse metric expressing the extent to which the system is free from hardware and software defects (or other defects) that can lead to system failures~\cite{zheng2014cloudqual}. Thus,
\begin{center}
    $R = 1 - FVT$. \\
\end{center}

The closer the Reliability value is to 1, the more reliable the system. $FVT$ represents the time-related flaws metric as previously defined.

\subsubsection{Functional Correctness.} The alternative reliability metric describes an error rate of the system in the sense of functional defects affecting the system processes and the procedures handling the data stored in the system~\cite{bures2018comprehensive}:
\begin{center}
    $FC = \frac{n_{failure} - n_{total}}{n_{total}}$, \\
\end{center}
where $n_{failure}$, and $n_{total}$, are the number of failed and total operations that have occurred in a time interval.

\subsubsection{Mean Time Between Failures.} To measure reliability of continuously running services, the mean time between failures (MTBF)~\cite{conte1986software,zheng2014cloudqual} can be used as follows:

\begin{center}
    $MTBF = \frac{\sum_{i=1}^{n}t_{i}-t_{i-1}}{n}$ \\
\end{center}

where $n$ is the number of detected failures in a set and $t_{i}$ is the (date) times of the individual SUT failures.


\subsubsection{Rate of User Error.}
User interaction with a system also plays an important role for the evaluation of its reliability. Thus, we can base a corresponding metric on the extent to which the user encounters errors or is required to perform an action arising from a system error:
\begin{center}
    $RUE = \frac{n_{user\_failure} - n_{total}}{n_{total}},$
\end{center}
where $n_{user\_failure}$  is the number of failed user operations and $n_{total}$ is the total number of user operations that have occurred in a time interval~\cite{sollie2005security}.

\subsubsection{Responsiveness.} To express the extent to which the system respond to the requests during a time interval, the responsiveness function suggested by Zheng~\textit{at al.} \cite{zheng2014cloudqual} can be used: 

\begin{center}
    $RESP = 1 - \frac{f_{i=1}^{n}(t_{i})}{t_{max}}$, 
\end{center}

where $t_{i}$ is the time between the submission and the completion of $i$th request, $t_{max}$ denotes the maximal acceptable time to complete a request, and the function $f$ is an abstraction for a function expressing the tendency of the observed data, e.g., the mean or median~\cite{zheng2014cloudqual}. A value of $RESP$ closer to 1 means better system responsiveness.

\subsubsection{Security.} For a high-level expression of the security of an IoT system, metrics based on \textit{Flaws over Time} can be employed:

\begin{center}
    $SEC = 1 - FVT_{sec}$, 
\end{center}

where $FVT_{sec}$ is the $FVT$ metric capturing the security flaws and defects. This metric ranges from 0 to 1, where 1 implies the highest security level. An alternative option can use the \textit{mean time between failures}. In this case, the definition of $MTBF$ can be maintained; only the failures taken into account are considered security breaches and incidents detected during the run time of the system.

To further analyse the security of an IoT system, Bonilla~\textit{et al.} suggest grouping the security flaws per IoT layer (i.e. perception, network and application) to assist in a deeper understanding~\cite{8328467}. The authors define the security of level $i$ as

\begin{center}
    $SEC_i = (\sum_{j=1}^{k_i}{V_{ij} \cdot T_{ij} \cdot E_{ij}) \cdot A_i},$
\end{center}

where $i$ is the IoT layer, $k_i$ is the number of known vulnerabilities in layer $i$, for each vulnerability $j$ in layer $i$, Search Results
Web result with site links Common Vulnerability Scoring System~(CVSS) base score~\cite{mell2006common}, the weight of the vulnerability class and the vulnerability exploitability factor are represented as $V_{ij}$, $T_{ij}$, and $E_{ij}$ respectively. Finally,  the authors define $A_i$ is the asset weight for layer $i$, which is determined based on its number of vulnerabilities~\cite{8328467}.

\subsection{Metrics to Evaluate Effectiveness of the Testing Process}
\label{sec:metrics_testing_process}

For the expression and measurement of the performance of the testing process of an IoT system, several metrics can be used. Compared to standard software development, in this area, established metrics can be reused, as discussed specificity of IoT systems does not play a more substantial role here.

\subsubsection{Test to Defect Ratio.}
To measure the quality of the IoT system, a metric based on a number of discovered defects per executed test case can be used:
\begin{center}
    $TDR = \frac{n_{defects}}{n_{steps}}$,
\end{center}

where $n_{defects}$ denotes the number of defects discovered in a defined time period (typically the testing phase of a test level) and $n_{steps}$ denote number of test steps in the test cases executed in a given time period. Alternatively, if the test management process does not allow for the tracking of the individual steps of the test cases, number of test cases must be used instead of $n_{steps}$. The high value of the $TDR$ indicates the high relative density of the defects in an actual version of the examined system.

\subsubsection{Test Execution Productivity.} To measure the time effectiveness of test execution, a metric for the amount of labor required to execute the test cases can be used~\cite{nirpal2011brief}:
\begin{center}
    $TEP = \frac{TS + TS_{retested}}{E}$,
\end{center}

where $TS$ denotes the number of total executed test steps in a given time period, $TS_{retested}$ denotes the number of additional retested steps in the given period, and $E$ denotes the labor required to conduct these test steps, measured in personnel hours.

\subsubsection{Defect Rejection Rate.}

Due to the poor reporting quality of the found defects, some of them were rejected by the development team, which caused additional overhead in the system development and testing process (such defects must be verified by a tester, reported again with an improved description, and re-analyzed by the development team). On large scale projects, the increase in overhead caused by such sub-optimal processes can be significant. The defect rejection rate is expressed as:

\begin{center}
    $DR = \frac{n_d}{n_{rd}} \cdot 100\%$
\end{center}

where $n_{rd}$ denotes the number of rejected defects, and $n_d$ denotes the number of total reported defects for a given period or module of the system~\cite{nirpal2011brief}. Alternatively, defect acceptance can be defined when $n_{rd}$ is substituted by the number of defects accepted by the development team to be fixed~\cite{nirpal2011brief}.

\subsubsection{Test Scripting Productivity.} The effectiveness of the test case creation in a test preparation phase (or during the testing) is expressed as:

\begin{center}
    $SP = \frac{n_{test\_steps}}{E}$,
\end{center}

where $n_{test\_steps}$ is the number of created test steps, and $E$ denotes the labor required to create them, measured in personnel hours. Alternatively, the number of test cases can be used instead of the number of test steps; however, because the test cases might differ in length and level of detail, the accuracy of such metric might be lower.

\subsubsection{Requirement Coverage.} To measure the extent to which the system functionality is covered by the test cases, the requirements gathered in the requirement phase should be mapped to those test cases. If such traceability~\cite{van2013tmap} is available, the coverage can be quantified as:

\begin{center}
    $RC = \frac{n_{mapped\_rq}}{n_{total\_rq}} \cdot 100\%$,
\end{center}

where $n_{mapped\_rq}$ denotes the requirements that are mapped to any test case and $n_{total\_rq}$ denotes the number of total requirements. This metric can serve as the main indicator of basic flaws in the test coverage; a value below $100\%$ generally requires further investigation. 

As each requirement can be covered by a set of test cases, more detailed test coverage metrics can be used to obtain more accurate insight into the test coverage. However, such an overview is out of the scope of our study, and we recommend further literature on this topic in Section~\ref{sec:discussion}.

\subsubsection{Defect Discovery vs Defect Fix Rate.}
To support managerial decisions regarding releasing an IoT system or establishing the transition between individual test levels, the speed at which the new defects can be repaired by the development team can be evaluated as:

\begin{center}
    $DD = \frac{dd_{c/h/m}}{dr_{c/h/m}}$,
\end{center}
where $dd_{c/h/m}$ is the number of defects discovered in the last $N$ days that are of critical, high, or medium severity, and $dr_{c/h/m}$ is the number of closed and rejected defects (defects that are considered fixed after proper retesting or rejected by a test manager or the development team) in the last $N$ days that are of critical, high, or medium severity.

A lower $DD$ value indicates a higher quality of the developed system, as well as, a better prospective capacity of the development team to fix the remaining defects. This metric can also be used separately for defects of different severity (critical, high, or medium) or the numbers of defects can be evaluated cumulatively.

\subsubsection{Test Execution Rate.} To monitor the test progress, the following metric can be used:

\begin{center}
    $TER = \frac{n_{tests\_not\_executed}}{n_{tests\_planned}} \cdot 100\%$,
\end{center}
where $n_{tests\_not\_executed}$ is the number of tests not executed and $n_{tests\_planned}$ is the total number of planned tests. The metric can be used for continuous monitoring of testing progress or for the evaluation of tests that were executed at the end of a test level \cite{nirpal2011brief}.

\subsubsection{Test Case Reuse in Regression Tests.}
For high-level quantification of the conducted regression tests, metrics expressing the reuse of test cases can be applied:

\begin{center}
    $TCR = \frac{n_{tc\_used\_in\_rt}}{n_{tc\_total}} \cdot 100\%$,
\end{center}

where $n_{tc\_used\_in\_rt}$ denotes the number of test cases used in the regression tests and $n_{tc\_total}$ denotes the number of all test cases created during the system creation.

A low value of $TCR$ indicates the inability to reuse previously created test cases in regression testing; hence, a low level of regression tests is probable, which might be sub-optimal from a test management viewpoint.

\subsubsection{Defect Re-open Rate.}

The effectiveness of the removal of defects from a system can be measured as:

\begin{center}
    $DRR = \frac{n_{reopened}}{n_{fixed}}$,
\end{center}

which is a ratio of inadequately fixed defects reopened during retesting ($n_{reopened}$) and fixed defects that have been successfully retested ($n_{fixed}$)~\cite{nirpal2011brief}. 

In practical terms, $DRR$ describes the quality of the defect fixes and the quality of the defect reporting, as vague defect reports might lead to weak defect fixes. The high value of $DRR$ indicates the ineffectiveness of the defect fixing process.

\subsubsection{Defect Density of Test Case Review.}

To ensure the quality of created test cases, their review is recommended during the test preparation phase. The review results indicating quality of created test cases can be expressed as:

\begin{center}
    $RDD = \frac{n_{rd}}{n_{rtsur}}$,
\end{center}

which is the ratio of total reviewed test case defects or flaws ($n_{rd}$) to the total of raw test steps under review ($t_{rtsur}$); the lower the $RDD$ value, the more efficient the test script creation team.

\subsection{Metrics Applicable to Both Previous Aspects}
\label{sec:metrics_universal}

Some metrics measure both the quality of the IoT products and the test process performance. In this section, we suggest that such metrics are applicable to IoT systems.

\subsubsection{Defect Leakage.}
To express the extent to which the defects are not detected in certain test levels and are discovered in the following test level (or in a production run of a system after its release), the defect leakage can be defined as: 

\begin{center}
    $DL = \frac{n_{i+1}}{n_{i}+n_{i+1}} \cdot 100\%$,
\end{center}
where $n_{i+1}$ is the number of valid defects detected in phase $i+1$, and $n_{i}$ is the number of valid defects detected in phase $i$. The phase $i+1$ can be considered as the production run of the system after its rollout.

In addition to expressing the actual state of the tested system, the $DL$ also indirectly indicates the effectiveness of the testing team and testing process, along with the quality of the created test cases.

There are alternative names or definitions of this metric in the literature. Nirpal~\textit{et al.} refer to this metric as test efficiency~\cite{nirpal2011brief} and Chen~\textit{et al.} provide a similar metric called~\textit{Test Effectiveness}~\cite{chen2004effective}, defined as:

\begin{center}
    $TE = \frac{n_{T}}{n_{TP}+n_{F}} \cdot 100\%$,
\end{center}

where $n_{T}$ is the number of defects detected during the product cycle, $n_{TP}$ is the number of defects detected during the test phases, and $n_{F}$ is the number of defects detected in the system in its production run~\cite{chen2004effective}.

\subsubsection{Effective Defect Density.}
As not all detected defects have the same significance, a metric using a weighted number of defects can be used for better reporting accuracy; this is defined as:

\begin{center}
    $EDD = \frac{n_{wd}}{n_{steps}}$,
\end{center}

where $n_{wd}$ can be computed as the mean of the number of defects weighted by their severity, and $n_{steps}$ is the number of test case steps \cite{nirpal2011brief}. 

For better accuracy, $EDD$ shall not be used for the system as a whole, as different parts of the system can contain a significantly different number of defects. Instead, $EDD$ shall be computed for individual system parts.

\subsubsection{Valid Defects.} As reported, the defects are exchanged between the testers and the development team, and some of the defects can be rejected by developers during the overall process (which can be expressed by the  \textit{Defect Rejection Rate} metric). The goal is to make the defect fixing process more effective and objective for all involved parties; hence, the review of reported defects must be conducted by a dedicated test manager or, in later test levels, a representative of prospective users from a user acceptance testing team. Such reviews can also eliminate the duplication of reported defects. A metric based on a ratio of valid defects can be used to support such a process, defined as: 

\begin{center}
    $VD = \frac{n_{valid\_defects}}{n_{total\_defects}}$,
\end{center}

where $n_{valid\_defects}$ denotes the number of relevant defects and $n_{total\_defects}$ is the total number of reported defects.

\subsubsection{Quality of Code.} To evaluate quality of a software part in an IoT system, a metric using the number of defects and the amount of newly added code can be used:

\begin{center}
    $QC = \frac{D_{TP}+D_{F}}{KCSI}$,
\end{center}

where $D_{TP}$ is the number of defects found in the testing phases, $D_{F}$ is the number of defects found in a production run of the system, and $KCSI$ is the number of new lines of source code or changed code in a development phase, scaled in thousands~\cite{chen2004effective}. Chen~\textit{et al.} suggest using $KCSI$ instead of the total number of code lines to emphasize actual quality achieved in a particular development phase~\cite{chen2004effective}.

\section{Discussion}
\label{sec:discussion}

In this study, we consolidated the available reliability and quality metrics applicable to IoT systems, as no previous attempt had been made for a high-level reliability view in this domain.

However, this list might not be exhaustive for several reasons. Because of the scope selection, we have not covered several specific areas or domains of metrics, which may be considered, even indirectly, relevant to the problem. 

Namely, we have not focused on the~\textit{QoS} field, as the QoS is measured in the lower levels of an IoT system and this area has already been covered in the literature~\cite{snigdh2016quality,8519862,white2017quality,ming2012modeling}. However, there are overlaps between the scope definition used in our study and the QoS field; typically,  the \textit{Availability} and \textit{Responsiveness} metrics could be influenced by the QoS on the lower levels of an IoT system.

In addition, it is up to debate, if \textit{test coverage} of the created test cases shall be discussed in the context of metrics related to a performance of the testing process. When properly applied, test coverage metrics generally provide a conception of the potential of created test cases to detect some defects. However, as a number of test coverage criteria have been defined for different types of tests and this problem has also been covered in the literature~\cite{pezze2008software,ammann2016introduction,ammann2008coverage,dias2007survey}, we decided to approach this problem using the \textit{Requirement Coverage} metric as it best fit the high-level viewpoint previously mentioned.

Moreover, another class of metrics that might be potentially relevant when discussing the quality and reliability of an IoT system are metrics related to \textit{Code Quality}. Poor code quality can negatively impact the effectiveness of the testing process and the quality and reliability of the created IoT system. However, to maintain viewpoint consistency and because this topic has been sufficiently discussed in the literature~\cite{baggen2012standardized,heitlager2007practical,pantiuchina2018improving,chaparro2014impact,jiang2008comparing}, we decided not to include specific code quality metrics in the presented overview. The only exception is the high-level metric \textit{Quality of Code} suggested by Chen \textit{et al.}~\cite{chen2004effective}, which we considered to fit well with the high-level framework used here.

Another area to be discussed is the application of the presented metrics. In particular, certain concerns can be raised regarding the unsuitable application of some metrics for evaluating the effectiveness of the testing process (see Section~\ref{sec:metrics_testing_process}). Generally, these metrics can be useful in such measurements; however, when they are applied to team key performance indicators (KPIs), individual performance evaluations, or individual KPIs, or contractual conditions, negative side effects are possible. 

When incorrectly applied to monitor the test preparation phase, the \textit{Test Scripting Productivity} metric can lead to a higher amount of brief test cases, which are quickly prepared without the necessary in-depth analysis of the particular situations to be tested. Logically, the potential to detect defects of such test cases might be lower as a consequence.  Hence, the use of such metrics should be balanced by test case revisions (that can be supported by \textit{Test Case Review Defect Density} metric) or the parallel application of another metric focusing on the defect detection capacity of created test cases, such as \textit{Test to Defect Ratio}.

Likewise, the overemphasis on \textit{Test Execution Productivity} may have an adverse effect, as the testers focus only on the exact scenario described in a test case and are discouraged from trying more data combinations or alternative situations, which could lead to the discovery of more relevant defects. This metric can be balanced, for example, by a total number of found relevant defects. 

Finally, a misplaced emphasis on the \textit{Defect Rejection Rate} may discourage testers from reporting defects, the relevance of which may be unclear. Thus, they may report only the defects in which the actual SUT functionality explicitly contradicts a scenario given in a test case, and valuable information about other potential defects might be systematically lost during the testing process. This metric can be balanced when used together with the total number of reported defects or a metric capturing the relevance of the found defects (\textit{Valid Defects}, as mentioned in Section \ref{sec:metrics_universal}).

\section{Conclusion}

In this paper, we analyzed the contemporary literature dedicated to the quality measurements and metrics for IoT systems and consolidated applicable metrics to a unified view. To our knowledge, such a view has not yet been proposed for IoT systems. 

In this unified view, we have taken a high-level perspective on the functionality, reliability, and related quality aspects of IoT systems. To maintain consistency with this overview, the QoS metrics and low-level code metrics were deliberately not examined here. These specific metrics are detailed in the existing literature, which we provided in Section \ref{sec:discussion}.

We divided the consolidated metrics into three categories: metrics for measuring system reliability and quality, metrics for evaluating the testing processes of IoT systems, and metrics that can be applied for both purposes.

The list of metrics provided here might not be exhaustive, as a wide variety of current IoT systems also imply their domain specificity, and we deliberately did not address specific lower-level fields such as QoS, test coverage, and code quality. However, we believe that we have covered the majority of the important high-level quality metrics of IoT systems, and the provided overview will be useful for both researchers and testing practitioners in the field.

\section*{Acknowledgements}

\textit{This research is conducted as a part of the project TACR TH02010296 Quality Assurance System for the Internet of Things Technology. The authors acknowledge the support of the OP VVV funded project CZ.02.1.01/0.0/0.0/16\_019 /0000765 “Research Center for Informatics”. Bestoun S. Ahmed has been supported by the Knowledge Foundation of Sweden (KKS) through the Synergi Project AIDA - A Holistic AI-driven Networking and Processing Framework for Industrial IoT (Rek:20200067).}

%
%
%
\bibliographystyle{splncs04}
\bibliography{paper}

\end{document}